\documentclass[superscriptaddress]{revtex4}
\usepackage{amsmath,amssymb,amsthm}
\newcommand{\al}{\alpha}
\newcommand{\be}{\beta}
\newcommand{\ga}{\gamma}
\newcommand{\ep}{\epsilon}
\newcommand{\la}{\lambda}
\newcommand{\om}{\omega}
\newcommand{\si}{\sigma}
\newcommand{\vp}{\varphi}
\newcommand{\ze}{\zeta}
\newcommand{\La}{\Lambda}
\def\x{\mathbf{x}}
\def\w{\mathbf{w}}
\def\z{\mathbf{z}}
\newcommand{\bze}{\boldsymbol{\ze}}
\def\BH{\,\overline{\!H}{}}
\newcommand{\cM}{{\mathcal M}}
\newcommand{\cN}{{\mathcal N}}
\newcommand{\cP}{{\mathcal P}}
\newcommand{\cR}{{\mathcal R}}
\newcommand{\cS}{{\mathcal S}}
\def\C{\mathbb{C}}
\def\N{\mathbb{N}}
\def\R{\mathbb{R}}
\def\RP{\mathbb{RP}}
\newcommand{\fS}{{\mathfrak S}}
\newcommand{\fsl}{{\mathfrak{sl}}}
\def\sn{{\rm sn}}
\def\cn{{\rm cn}}
\def\dn{{\rm dn}}

\newcommand{\myspan}{\operatorname{span}}
\newcommand{\diff}{\operatorname{d}\!}
\newcommand{\iu}{{\rm i}}
\newcommand{\e}{{\rm e}}
\newcommand{\pa}{\partial}
\newcommand{\ra}{\rightarrow}
\newcommand{\id}{1\hspace{-.25em}{\rm I}}
\def\ni{\noindent}
\newtheorem{theorem}{\bf Theorem}
\newtheorem{lemma}{\bf Lemma}
\newtheorem{corollary}{\bf Corollary}
\newtheorem{proposition}{\bf Proposition}
\newcounter{rem}
\newenvironment{remark}{{\ni\em Remark~\therem.}}{\newline\stepcounter{rem}}
\begin{document}
\title{$A_N$-type Dunkl operators and new spin
Calogero--Sutherland models}
\author{F. Finkel}
\author{D. G\'omez-Ullate}
\author{A. Gonz\'alez-L\'opez}
\author{M. A. Rodr\'{\i}guez}
\affiliation{Departamento de F\'{\i}sica Te\'orica II,
Universidad Complutense,
28040 Madrid, Spain}
\author{R. Zhdanov}
\altaffiliation{On leave of absence from Institute of Mathematics,
3 Tereshchenkivska St., 01601 Kyiv - 4 Ukraine}
\affiliation{Departamento de F\'{\i}sica Te\'orica II,
Universidad Complutense,
28040 Madrid, Spain}
\date{January 26, 2001}
\begin{abstract}
A new family of $A_N$-type Dunkl operators
preserving a polynomial subspace of finite dimension
is constructed. Using a general quadratic combination
of these operators and the usual Dunkl operators,
several new families of exactly and quasi-exactly
solvable quantum spin Calogero--Sutherland models are obtained.
These include, in particular, three families of
quasi-exactly solvable elliptic spin Hamiltonians.
\end{abstract}
\maketitle

\section{Introduction}\label{intro}
In the early seventies, Calogero~\cite{calo71} and
Sutherland~\cite{suther71,suther72} introduced the celebrated
exactly solvable (ES) and integrable quantum many-body problems in one
dimension that bear their names. These papers had a profound
impact in the whole physics community, as reflected by the vast
amount of literature devoted to the study of the mathematical
properties and applications of these models. Among the most recent ones we could
mention soliton theory~\cite{Ka95,Po95}, orthogonal
polynomials~\cite{LV96,BF97,dunkl98}, fractional statistics and
anyons~\cite{CL99}, random matrix theory~\cite{TSA95}, and
Yang--Mills theories~\cite{GN94,DF98}, to name only a few. Later
on, Olshanetsky and Perelomov~\cite{OP83} explained the
integrability of the original Calogero--Sutherland (CS) models by
relating them to the root system of $A_N$ type. These
authors then constructed new families of integrable many-body
Hamiltonians associated with all the other root systems. Furthermore,
they showed that the most general interaction
potential for these models is proportional to the Weierstrass
$\wp$ function.

A considerable effort has been devoted over the last decade to the
extension of CS models to particles with spin. These models are a
step forward towards the unification of the CS scalar models and
integrable spin chains, like the Haldane--Shastry
model~\cite{Ha88,Sh88}. Several different techniques have been
used to construct spin counterparts of the scalar CS models,
including the exchange operator method~\cite{poly92}, the Dunkl
operators formalism~\cite{ber93,cher94}, the supersymmetric
approach~\cite{tur98}, and reduction by discrete
symmetries~\cite{poly99a}. In the
quantum case, only the rational and trigonometric (or hyperbolic)
spin CS models have been constructed, both in their
$A_N$~\cite{poly92,hikami92,ber93,brink93,basu96,In97,take97,UNW98} and
$BC_N$~\cite{yam95} versions.
In the $A_N$ case, the integrability and the exact-solvability
of these models both follow from the fact that
the Hamiltonian is related to a quadratic combination of
some family of Dunkl operators.

The Dunkl operators
\begin{equation}\label{r2}
T_i={\partial\over \partial z_i} + {a} \sum\limits_{j\ne i}\,
{1\over z_i-z_j}(1-K_{ij}), \qquad i=1,\dots,N\,,
\end{equation}
were originally introduced in~\cite{dunk89}
in connection with the theory of orthogonal polynomials associated
with finite reflection groups. In the latter expression,
$a$ is an arbitrary real parameter and the sum runs over
$j=1,\dots,i-1,i+1,\dots,N$. The permutation operators
$K_{ij}=K_{ji}$ act on an arbitrary function $f(\z)$, with $\z
=(z_1,\ldots, z_N) \in \R^N$, as
\begin{equation}\label{KsAction}
(K_{ij}f)\,(z_1,\ldots, z_i,\ldots, z_j,\ldots,z_N) =
f(z_1,\ldots, z_j,\ldots, z_i,\ldots,z_N).
\end{equation}
Using the relations
\begin{equation}\label{KsAlgebra}
K_{ij}^2=1,\quad K_{ij}K_{jk}=K_{ik}K_{ij}=K_{jk}K_{ik},\quad
K_{ij}K_{kl}=K_{kl}K_{ij},
\end{equation}
where $i, j, k, l$ take different values in the range
$1,\dots,N$, one can establish the commutativity of the
operators~\eqref{r2} and prove that $T_i,
K_{jk}$,\ $i,j,k = 1,\ldots, N$, span a realization of a
degenerate Hecke algebra (see \cite{cher94} for more details).
Since the rational spin CS Hamiltonian is related to a
polynomial in the Dunkl operators~\eqref{r2}, these
operators yield a complete set of commuting integrals
of motion. In addition, the spectrum of the Hamiltonian
follows immediately from that of the Dunkl operators, which
can be easily computed~\cite{ber93,basu96}.

The previous considerations also apply to the operators
\begin{equation}\label{r01}
\tilde T_i= z_i{\partial\over \partial z_i} + {a} \sum\limits_{j<
i}\, {z_i\over z_i-z_j}\,(1-K_{ij})+ {a} \sum\limits_{j> i}\,
{z_j\over z_i-z_j}\,(1- K_{ij})+1-i ,
\end{equation}
$i=1,\dots,N$, introduced by Cherednik~\cite{cher94}
in connection with the trigonometric spin CS model. In other
words, the operators $\tilde T_i$ commute,
have an easily computable spectrum, and can be used
to obtain a complete set of integrals of motion
and the spectrum of the Hamiltonian. It
has become customary in the literature to refer to both families
of operators $T_i$ and $\tilde T_i$ as Dunkl operators.

Recently, some partially solvable deformations of the scalar CS models
with an external potential have been proposed~\cite{MRT96,HS99,GGR00}.
For these models ---in contrast to the CS models listed in~\cite{OP83}---
only a finite-dimensional subset of the spectrum can be computed
algebraically. Following Turbiner and Ushveridze~\cite{TU87,Tu88}, we shall
use the term {\em quasi-exactly solvable} (QES)
to refer to this type of models; see also the
reviews~\cite{Sh89,GKO94,Us94}. In all these models, the Hamiltonian
can be expressed as a quadratic combination of the generators
of a realization of $\fsl(N+1)$ by
first-order differential operators preserving a finite-dimensional
space of smooth functions. The action of the
Hamiltonian in this space can thus be represented by a finite-dimensional
constant matrix. The eigenvalues of this matrix
belong to the spectrum of the Hamiltonian provided the
corresponding eigenfunctions satisfy some appropriate boundary conditions.
The Lie algebra $\fsl(N+1)$ is usually referred to as a
{\em hidden symmetry algebra} of the Hamiltonian in these models.

In this paper, we propose a general procedure for constructing
(Q)ES spin CS models, close in spirit to the
hidden symmetry algebra approach to scalar QES models.
The starting point of our construction is the well-known fact that
the two standard families of Dunkl operators~\eqref{r2}
and~\eqref{r01} admit an infinite
sequence of invariant polynomial subspaces of finite dimension.
One of the main novelties in our approach consists in
the introduction of a new family of commuting Dunkl operators
which, together with the other two families~\eqref{r2} and~\eqref{r01},
is shown to preserve a single polynomial subspace of finite dimension.
We then prove that certain quadratic combinations
involving {\em all} three families of Dunkl operators always
yield a spin CS Hamiltonian. The
QES character of these Hamiltonians follows immediately from
the fact that the Dunkl operators admit a finite-dimensional
invariant subspace. Moreover, if the original quadratic
combination does not involve the new family of Dunkl operators,
the resulting Hamiltonian preserves an infinite sequence of
finite-dimensional subspaces of smooth functions,
which we shall take as the definition of exact solvability.
The linear space spanned by all three types of Dunkl operators
is then shown to be invariant under the projective
action of the group ${\rm GL}(2,\R)$.
We make use of this fact to perform a complete classification
of the resulting (Q)ES spin CS models.
All the previously known exactly-solvable spin CS models of $A_N$ type
appear as particular cases, arising from a quadratic combination of a
{\em single} type (either~\eqref{r2} or~\eqref{r01}) of Dunkl operators.
In addition, we obtain many new spin CS models, both exactly
and quasi-exactly solvable. These include, in particular, several
elliptic QES spin CS models. To the best of our knowledge,
these are the first examples of solvable quantum spin CS models involving
elliptic functions.

\section{A new family of $A_N$-type Dunkl operators }\label{sec.dunkl}

In this section we shall define a third family of Dunkl operators
that preserve certain finite-dimensional polynomial subspaces.
These three families shall be used in the following sections to
construct exactly and quasi-exactly solvable spin many-body
Hamiltonians.

Let us begin by introducing the polynomial subspaces
\begin{align}
\mathcal R_m(\z) &= \text{\rm span} \left\{ \prod\limits_{i=1}^N\,
z_i^{l_i}\;:\; l_i \le m,\quad i=1,\ldots, N \right\}\,, \\
\mathcal P_n(\z) &= \text{\rm span} \left\{ \prod\limits_{i=1}^N\,
z_i^{l_i}\;:\; \sum_{i=1}^N l_i \le n \right\}\,,
\end{align}
which shall be referred to as the \emph{rectangular} and
\emph{triangular} modules, respectively, by analogy with the two
particle case \cite{FK98}.

A well-known property of the Dunkl operators~\eqref{r2}, \eqref{r01}
is the fact that they preserve the
triangular module $\mathcal P_n(\z)$ for arbitrary $n$. Seemingly
less known, but central to our construction, is the fact that they
preserve the rectangular module $\mathcal R_m(\z)$ for arbitrary
$m$ as well. On the other hand, the differential parts of the
Dunkl operators (\ref{r2}), (\ref{r01}) together with the
differential operator $z_i^2\partial_{z_i}$, span a realization of
$\mathfrak{sl}(2)$. Inspired by this fact, it is natural to
suggest the following ansatz for a third set of Dunkl operators:
\[
J_i=z_i^2{\partial\over \partial z_i} - mz_i + \sum\limits_{j\ne
i}\, f_{ij}(\textbf{z})(1-K_{ij}),\qquad i=1,\dots,N,
\]
where $m$ is an arbitrary non-negative integer and $f_{ij}(\z)$ is
a function anti-symmetric in $i, j$. This new family does not
preserve the module $\mathcal P_n(\z)$, but for a suitable choice
of the functions $f_{ij}(\z)$ it will be shown to preserve the
module $\mathcal R_m(\z)$.

To this end, let us define the operators
\begin{equation}\label{Qs}
\begin{aligned}
& Q_i^-= {a} \sum\limits_{j\ne i}\,{1\over z_i-z_j}\,(1-K_{ij}),\\
& Q_i^0= \frac{{a}}{2}\sum\limits_{j\ne i}\, {z_i+z_j\over z_i-z_j}\,
(1-K_{ij}), \\
& Q_i^+={a}\sum\limits_{j\ne i}\, {z_iz_j\over z_i-z_j}\,(1-K_{ij}),
\end{aligned}
\end{equation}
where $a$ is a real parameter and the sum runs over $j=1,\dots
i-1, i+1,\dots N$.

The following lemma will be important in the sequel:
\begin{lemma}\label{QLemma}
For any non-negative integer $n$, the rectangular module $\mathcal
R_n(\z)$ is invariant under the action of the operators $Q_i^-$,
$Q_i^0$ and $Q_i^+$. The triangular module $\mathcal P_n(\z)$ is
invariant only under the action of the operators $Q_i^-$ and
$Q_i^0$.
\end{lemma}
\begin{proof}
It suffices to prove that the inclusions
\begin{align*}
&{ h^\ep_{ij} \over z_i-z_j}( 1-K_{ij} ) \;\mathcal R_n(\z)
\subset \mathcal R_n(\z) \,, \quad \ep=\pm,0,\\
&{ h^\ep_{ij} \over z_i-z_j}( 1-K_{ij} ) \;\mathcal P_n(\z)
\subset \mathcal P_n(\z)  \,, \quad \ep=-,0,
\end{align*}
hold for any pair of indices $i\ne j$, where
\[
h^-_{ij} = 1 \,,\qquad h^0_{ij} = z_i+z_j \,,\qquad h^+_{ij} =
z_iz_j.
\]
The action of these operators on an arbitrary monomial
$$
{h^\ep_{ij}\over z_i-z_j}\,(1-K_{ij})\prod_{k=1}^N
z_k^{l_k}= {z_i^{l_i} z_j^{l_j}-z_i^{l_j} z_j^{l_i}\over
z_i-z_j}\,{h^\ep_{ij}\over z_i^{l_i} z_j^{l_j}}\prod_{k=1}^N
z_k^{l_k}
$$
yields a polynomial, since $z_i^{l_i} z_j^{l_j}-z_i^{l_j}
z_j^{l_i}$ is a multiple of $z_i-z_j$. The homogeneous degree of
this polynomial is either $0$ (if $l_i= l_j$) or $ \ep+
\sum_i l_i $ (if $l_i\ne l_j$). Therefore, if the original
monomial belongs to $\mathcal P_n(\z)$ the resulting polynomial
is also in $\mathcal P_n(\z)$ for $\ep=-,0$, but lies outside
$\mathcal P_n(\z)$ for $\ep=+$.

On the other hand, the degrees of the variables $z_k$ in the
resulting polynomial remain equal to $l_k$ if $k \ne i,j$, while
the degrees of $z_i$ and $z_j$ satisfy
\[
\deg(z_k) \leq \max(l_i,l_j)-1+d_\ep\,,\quad k=i,j\,,
\]
where
\[
d_\ep =
  \begin{cases}
    0 & \text{if $\ep=-$}, \\
    1 & \text{if $\ep=0,+$}.
  \end{cases}
\]
Therefore, if the original monomial belongs to the space $\mathcal
R_n(\z)$ so does the resulting polynomial in all three cases
$\ep=\pm,0$. \qed
\end{proof}
The following three sets of Dunkl operators shall be the building
blocks for the construction of several new (quasi-)exactly
solvable spin CS models:
\begin{equation}
\begin{aligned}
& J_i^-={\partial\over \partial z_i} + {a} \sum\limits_{j\ne
i}\,
{1\over z_i-z_j}(1-K_{ij}),\\
& J_i^0= z_i{\partial\over \partial z_i} - \frac{m}{2} +
\frac{{a}}{2} \sum\limits_{j\ne i}\, {z_i+z_j\over z_i-z_j}
(1-K_{ij}), \\
& J_i^+= z_i^2{\partial\over \partial z_i} - mz_i + {a}
\sum\limits_{j\ne i}\, {z_iz_j\over z_i-z_j}(1-K_{ij}),
\end{aligned}\label{Js}
\end{equation}
where $i=1,\dots,N$, $a$ is a real parameter, and $m$ is a
non-negative integer.
Note that the operators $J_i^-$ coincide exactly with the Dunkl
operators \eqref{r2}, while the operators $J_i^0$ differ from the
Cherednik operators \eqref{r01} by a linear combination with
constant coefficients of the permutation operators $K_{ij}$,
namely
\begin{equation}\label{relation}
\tilde T_i = J_i^0+\frac a2\sum_{j<i}(1-K_{ij})
-\frac a2\sum_{j>i}(1-K_{ij})+\frac m2+1-i\,.
\end{equation}
The operators $J_i^+$ are, to the best of our knowledge, new.

The operators $J_i^\ep$ and $K_{ij}$ obey the following
commutation relations
\begin{align}
[J_i^{\pm},J_j^{\pm}]&=0\,, &
[J_i^0,J_j^0]&=\frac{a^2}4\sum_{k\neq i,j}K_{ij}
(K_{jk}-K_{ik})\,,\label{commJ}\\
[K_{ij},J_k^\ep]&=0\,,& K_{ij}J_i^\ep&=J_j^\ep
K_{ij}\,,\label{commJKR}
\end{align}
where $\ep=\pm,0$ and the indices $i,j,k$ are all different.
The set of operators
$$\left\{J_i^\ep,K_{ij}\;:\;i,j=1,\dots,N\right\},\qquad\ep=\pm,0, $$
spans a degenerate affine Hecke algebra, see~\cite{cher94}. This
is clear for $\ep=\pm$, while for $\ep=0$, it follows
from \eqref{relation} and the commutativity of the Cherednik
operators \eqref{r01}.

The key property in our construction of (quasi-)exactly solvable
spin CS models is the fact that the operators \eqref{Js} possess
invariant polynomial subspaces.
\begin{theorem}\label{QEStheorem}
The operators $J_i^-$ and  $J_i^0$  preserve the modules
${\mathcal P}_{n}(\z)$ and ${\mathcal R}_{n}(\z)$ for an arbitrary
non-negative integer $n$. The operators $J_i^+$ preserve the
module ${\mathcal R}_{m}(\z)$, but do not preserve the modules
${\mathcal P}_{n}(\z)$ and ${\mathcal R}_{k}(\z)$ for $k\ne m$.
\end{theorem}
\begin{proof}
The statement follows from Lemma~\ref{QLemma} and the fact that
the differential parts of $J_i^-$ and $J_j^0$ preserve the
modules ${\mathcal P}_{n}(\z)$ and ${\mathcal R}_{n}(\z)$ for any
 non-negative integer $n$, whereas the differential part
of $J_i^+$  preserves the module ${\mathcal R}_{k}(\z)$ only for
$k=m$. \qed
\end{proof}
The following corollary is an immediate consequence of
Theorem~\ref{QEStheorem}:
\begin{corollary}
Any polynomial in the operators $J_i^\ep$ leaves invariant
the rectangular module ${\mathcal R}_{m}(\z)$. In addition, if
the polynomial does not depend on $J_i^+$, it preserves the
modules ${\mathcal R}_{n}(\z)$ and ${\mathcal P}_{n}(\z)$ for all
$n$.
\end{corollary}

\section{Construction of spin Calogero--Sutherland models}\label{sec.cons}

In the previous section we have introduced a new set of Dunkl operators
preserving the space of polynomials $\cR_m$. Here we shall
make use of all three sets of Dunkl operators~\eqref{Js}
to construct some multi-parameter families of spin
CS models.

Consider the spin permutation operators $S_{ij}$, $i,j=1,\dots,N$,
whose action on a spin state $|s_1,\dots,s_N\rangle$, $-M\leq
s_i\leq M$, with $M\in\frac12\N$, is given by
\begin{equation}
S_{ij}|s_1,\dots,s_i,\dots,s_j,\dots,s_N\rangle=|s_1,\dots,
s_j,\dots,s_i,\dots,s_N\rangle\,.
\label{Sij}
\end{equation}
Note that the operators $S_{ij}$ obey the identities
\eqref{KsAlgebra} with $K_{ij}$ replaced by $S_{ij}$.
Let $\fS$ denote the linear space
$\myspan\big\{\,|s_1,\dots,s_N\rangle\big\}_{-M\leq s_i\leq M}$.
The action of the operators $S_{ij}$ in $\fS$ is thus represented
by $(2M+1)^N$-dimensional symmetric matrices.

The starting point of our procedure is the following quadratic
combination of the Dunkl operators~\eqref{Js}:
\begin{equation}\label{Jcombination}
\begin{aligned}
- H^* = \sum_i\Big( & c_{++}(J_i^+)^2 + c_{00}
(J_i^0)^2 +c_{--}(J_i^-)^2
+ \frac{c_{0+}}2\{J_i^0,J_i^+\}
+\frac{c_{0-}}2\{J_i^0,J_i^-\}\\
&+c_{+}J_i^+ +c_{0}J_i^0+c_{-}  J_i^-\Big),
\end{aligned}
\end{equation}
where $c_{\ep\ep'}, c_{\ep}\,,\; \ep,\ep'=\pm,0\,,$ are arbitrary
real constants. The term $\frac12\sum_i\{J_i^-,J_i^+\}$ differs
from $\sum_i(J_i^0)^2$ by a constant operator (see Appendix), and
for this reason it has not been included in~\eqref{Jcombination}.
We emphasize that only the particular cases
of~\eqref{Jcombination}
$$
-H^* =  c_{00}\sum_i(J_i^0)^2\,,\qquad
-H^* =  c_{--}\sum_i(J_i^-)^2
$$
have been previously discussed in the literature in connection with
CS models; see~\cite{poly92,ber93,cher94,yam95,dunkl98} and references therein.

As it is customary, we shall identify $K_{ij}$, $S_{ij}$, and $H^*$
with their natural extensions $K_{ij}\otimes\id$,
$\id\otimes S_{ij}$, and $H^\ast\otimes\id$ to the tensor
product $\C[z_1,\dots,z_N]\otimes\fS$.
The following lemma is an immediate consequence of Eq.~\eqref{commJKR}.
\begin{lemma}\label{commLemma}
The (Q)ES differential-difference operator $H^*$
commutes with $K_{ij}$ and $S_{ij}$ for all $i,j=1,\dots,N$.
\end{lemma}
This property plays a crucial role in the construction of
spin CS models; see, for instance, Ref.~\cite{basu96}.

Let $\La$ be the projection operator on states antisymmetric under
the simultaneous interchange of any two particles' coordinates and
spins. In terms of the total permutation operators $\Pi_{ij}=K_{ij}S_{ij}$,
the operator $\La$ can be alternatively defined by the relations
$\Pi_{ij}\La=-\La$, $j>i=1,\ldots,N$.
Since $K_{ij}^2=1$, these relations are equivalent to
\begin{equation}\label{La}
  K_{ij}\La=-S_{ij}\La,\qquad j>i=1,\ldots,N.
\end{equation}
For the lowest values of $N$ the antisymmetrizer $\La$ is given by
\begin{align*}
& N=2:\qquad \La=1-\Pi_{12}\,,\\
& N=3:\qquad \La=1-\Pi_{12}-\Pi_{13}-\Pi_{23}+\Pi_{12}\Pi_{13}+
\Pi_{12}\Pi_{23}\,.
\end{align*}
In general, $\La$ is an $(N-1)$-th degree polynomial in the
total permutation operators $\Pi_{ij}$. It thus follows from
Lemma~\ref{commLemma} that $H^*$ commutes with $\La$.

Suppose that $f(\z)$ is an eigenfunction of $H^\ast$ with eigenvalue $\la$.
For instance, $f$ could be one of the polynomial eigenfunctions that
$H^*$ is guaranteed to possess in $\cR_m$.
Given any (constant) spin state $|\si\rangle\in\fS$, the spin function
$\vp=\La [f(\z)|\si\rangle]$ is also an eigenfunction
of $H^*$ with the same eigenvalue $\la$.

Next, we introduce the matrix differential operator
$\BH$ obtained from $H^\ast$ by the formal substitutions
$K_{ij}\ra-S_{ij}$, $i,j=1,\dots,N$. The relations~\eqref{La}
imply that $\vp$ is a spin eigenfunction of $\BH$ with
eigenvalue~$\la$.

Using the formulae~\eqref{A1}--\eqref{A6} for the sums of the squares
and the anticommutators of the Dunkl operators~\eqref{Js} given
in the Appendix, we obtain the following explicit expression
for $\BH$:
\begin{equation}
\begin{aligned}\label{gaugeham}
-\BH &=\sum_i \left(P(z_i) \partial_{z_i}^2 + \tilde Q
(z_i)
\partial_{z_i} + R(z_i) \right)+2a c_{++}(1-m) \sum_{i<j} z_i z_j\\
&\quad + 2 a \sum_{i<j} \frac{1}{z_i-z_j} \left(
P(z_i)\partial_{z_i} -P(z_j)\partial_{z_j}\right) - a \sum_{i<j}
{P(z_i) + P(z_j) \over (z_i-z_j)^2} \left(1+S_{ij} \right)\\
&\quad+a \sum_{i<j} \left( c_{++} (z_i+z_j)^2+ c_{0+}
(z_i+z_j) + c_{00} \right) S_{ij} + \frac{a^2}{12}\,c_{00}
{\sum_{i,j,k}}'\!\left( 1 - S_{ij}S_{ik} \right)\!,
\end{aligned}
\end{equation}
where
\begin{align}
P(z) &= c_{++} z^4 + c_{0+} z^3 + c_{00} z^2+ c_{0-} z + c_{--} \,,\notag\\
\tilde Q(z) &= Q(z) + \Big(1-\frac{b}2\Big)\,P'(z)\,,\notag\\
Q(z) &= c_+ z^2 + c_0 z + c_- \,,
\qquad b=1+m+a(N-1)\,,\label{Q}\\
R(z) &= c_{++}\big(b+m(m-2)-1\big) z^2\!+\!\bigg[c_{0+} \Big[
\Big(1-\frac m2\Big)b+m(m-1)-1\Big]-m c_+ \bigg]z\notag\\
&\quad +\frac{c_{00}}4\,\big(2(b-1)+m(m-2)\big)
-\frac{m}2\,c_0\,,\notag
\end{align}
and ${\sum\limits_{i,j,k}}'$ denotes summation
in $i,j,k$ with $i\neq j\neq k\neq i$.

In the final step, one performs a gauge transformation with a suitable
scalar function $\mu(\z)$, followed by a change of variables
$\z=\bze(\x)$, $\x=(x_1,\dots,x_N)$,
\begin{equation}\label{gcv}
H=\mu\,\BH\,\mu^{-1}\Big|_{\z=\bze(\x)}\,,
\end{equation}
in order to reduce the {\em gauge spin Hamiltonian} $\BH$ to
the {\em Schr\"odinger form}
\begin{equation}\label{H}
H=-\sum_i \pa_{x_i}^2+V(\x)\,,
\end{equation}
where $V(\x)$ is a Hermitian matrix-valued function.
\begin{proposition}
The operator $\BH$ in Eq.~\eqref{gaugeham} can be reduced
to a matrix Schr\"o\-ding\-er operator~\eqref{H} by a change of
variables $\z=\bze(\x)$ and conjugation
by a scalar gauge factor $\mu(\z)$.
\end{proposition}
\begin{proof}
The gauge transformation~\eqref{gcv} with gauge factor
\begin{equation}\label{gaugefactor}
\mu(\z) =
\prod\limits_{i<j}\,(z_i-z_j)^{a}\,\prod\limits_{i=1}^N\,
P(z_i)^{-\frac{1}{4}}\, \exp\int^{z_i}\!
\frac{\tilde Q(y)}{2P(y)}\,\diff y,
\end{equation}
together with the change of variables
\begin{equation}
x_i=\ze^{-1}(z_i)=\int^{z_i}\!\frac{\diff y}{\sqrt{P(y)}}\,,\qquad
i=1,\ldots,N,\label{r8}
\end{equation}
map the gauge spin Hamiltonian $\BH$ to a matrix
Schr\"odinger operator~\eqref{H}, with potential
\begin{align}
V(\x) &= a^2{\sum_{i,j,k}}' {P(z_i) \over
(z_i-z_j)(z_i-z_k) }+ a \sum_{i \ne j}
{\tilde Q(z_i) \over z_i-z_j }+
a\sum_{i<j}{P(z_i) + P(z_j) \over
(z_i-z_j)^2}\,(a+S_{ij})\notag\\
&\quad -2a c_{++}(1-m) \sum_{i<j} z_i z_j
-a \sum_{i<j} \left( c_{++} (z_i+z_j)^2+ c_{+0} (z_i+z_j) +
c_{00} \right) S_{ij}\notag\\
&\quad+\frac14\sum_i \bigg[\frac{1}{P(z_i)}\bigg(\frac{3}4 P'(z_i)^2+\tilde
Q(z_i)^2-2\tilde Q(z_i) P'(z_i) \bigg)-P''(z_i)+2\tilde Q'(z_i)\notag\\
& \quad-4R(z_i)\bigg]-\frac{a^2}{12}\,c_{00}
{\sum_{i,j,k}}'\left( 1 - S_{ij}S_{ik} \right)\Bigg|_{\z=\bze(\x)}.
\qquad\qed\label{pot}
\end{align}
\end{proof}

\begin{remark}
The gauge factor $\mu(\z)$ in~\eqref{gaugefactor} was introduced
in~\cite{GGR00} in connection with a generalization of the theory
of QES models with one degree of freedom to many-body problems.
The existence of a (matrix or scalar) gauge factor and change of
coordinates reducing a given matrix second-order differential
operator in $N>1$ variables to a matrix Schr\"odinger
operator~\eqref{H} is not guaranteed a priori. In the scalar case
this problem was first addressed by Cotton~\cite{Co00}, while the
matrix case has been studied recently in Ref.~\cite{FK97}. In
fact, the quadratic combination $H^\ast$ has been chosen so that
a scalar gauge factor and change of variables can be easily found
for $\BH$. For instance, the term $\sum_i[J_i^+,J_i^-]$ has been
discarded because it involves first-order derivatives with
matrix-valued coefficients, which are usually very difficult to
gauge away.
\end{remark}

\begin{remark}\label{rem.trans}
The change of variables~\eqref{r8}, and hence the potential $V(\x)$,
are defined up to an arbitrary translation for each coordinate
$x_i$, $i=1,\dots,N$. We shall see that this arbitrariness
can be removed in some cases by requiring the potential to be
invariant under sign reversals of any coordinate $x_i$.
\end{remark}

If $\vp(\z)$ is one of the eigenfunctions of $\BH$ with
eigenvalue $\la$ constructed above, the spin function
$\psi(\x)=\mu\big(\bze(\x)\big)\vp\big(\bze(\x)\big)$
is clearly an eigenfunction of $H$ with the same eigenvalue.
Note, however, that we have not imposed so far any boundary
conditions on the eigenfunctions $\psi$. In general, the
parameters $a$, $c_{\ep\ep'}$, and $c_{\ep}$ defining $H^*$ should
satisfy certain constraints in order to ensure that the
appropriate boundary conditions are satisfied.

The following proposition is an immediate consequence of
the previous considerations:
\begin{proposition}
The spin Schr\"odinger operator~\eqref{H} with potential~\eqref{pot}
leaves invariant the module
\begin{equation}\label{M}
\cM_m=\mu\big(\bze(\x)\big)
\La\Big(\cR_m\big(\bze(\x)\big)\otimes\fS\Big).
\end{equation}
In addition, if $c_{++}=c_{0+}=c_+=0$, it preserves the modules $\cM_n$
and
\begin{equation}\label{N}
\cN_n=\mu\big(\bze(\x)\big)
\La\Big(\cP_n\big(\bze(\x)\big)\otimes\fS\Big),
\end{equation}
for any non-negative integer $n$.
\end{proposition}

If we add a constant term
\begin{equation}\label{V0}
V_0=\ga_0+\ga_1\sum_{i<j}S_{ij}+\ga_2{\sum_{i,j,k}}'S_{ij}S_{ik}\,,\qquad
\ga_i\in\R,
\end{equation}
to the potential~\eqref{pot}, the previous
procedure for constructing eigenfunctions of $H$ still applies
to $H+V_0$. Indeed, the associated operator $(H+V_0)^*$
is obtained by adding the term
$$
V_0^*=\ga_0-\ga_1\sum_{i<j}K_{ij}+\ga_2{\sum_{i,j,k}}'K_{ij}K_{ik}
$$
to the initial operator $H^*$. Our assertion follows from the fact
that $V_0^*$ preserves the modules
$\cP_n$ and $\cR_n$ for all $n$, commutes with all the permutation
operators $K_{ij}$, and acts trivially on $\fS$.
This observation will be used in what follows to simplify the formula
for $V(\x)$ by dropping terms of the form~\eqref{V0}.

\section{Classification of spin Calogero--Sutherland
models}\label{sec.class}

We have seen in Section~\ref{sec.cons} that any quadratic combination
of the form~\eqref{Jcombination} yields a spin CS model
for which a number of eigenvalues and their corresponding
eigenfunctions can be computed in an algebraic fashion. In
this section we shall obtain a complete classification of
the potentials constructed in this way.

The form of the potential $V(\x)$ in Eq.~\eqref{pot} depends on
the choice of parameters $c_{\ep\ep'}$ and $c_{\ep}$, $\ep,\ep'=\pm,0$.
The parameters $c_{\ep\ep'}$ which define the polynomial $P$ are
of particular significance, since they determine the form of the
change of variables~\eqref{r8}. However, different sets of
parameters $c_{\ep\ep'}$, $c_{\ep}$ defining the operator $H^*$
may give rise to the same potential. Indeed, there is a group of
residual transformations preserving the vector spaces
$\myspan\{J_i^-,J_i^0,J_i^+\}$, $i=1,\dots,N$. The image of the
operator $H^*$ under these transformations is still of the
form~\eqref{Jcombination}, albeit with different coefficients
$\hat c_{\ep\ep'}$ and $\hat c_{\ep}$. We shall make use of this
fact to classify the starting multi-parameter family of
operators~\eqref{Jcombination} into conjugacy classes. This will
provide a complete classification of the spin CS models
obtainable within our framework. The ideas used for this
classification are similar in spirit to those applied in
Refs.~\cite{GKO93,GKO94} to classify one-particle
Lie-algebraic QES Schr\"odinger operators.

Consider the mapping
\begin{equation}\label{Jconjugation}
J_i^\ep(\w)\,\mapsto
\widehat J_i^\ep(\z)=\mu_m(\z)\, J_i^\ep\big(\w(\z)\big)
\,\mu_m^{-1}(\z)\,,
\end{equation}
where $\w=(w_1,\dots,w_N)$ is given by the projective action
of ${\rm GL}(2,\R)$ on $\RP^1$ (M\"obius transformation)
\begin{equation}\label{moebius}
w_i={\alpha z_i + \beta\over \gamma z_i +
\delta}\;,\qquad i=1,\dots,N,\quad
\Delta=\alpha \delta - \beta \gamma \ne 0,
\end{equation}
and the gauge factor $\mu_m(\z)$ is defined by
\begin{equation}\label{gauge}
\mu_m(\z) = \prod_{i=1}^N (\gamma \,z_i + \delta)^m.
\end{equation}
The following lemma is an immediate consequence of the definition of the
Dunkl operators~\eqref{Js}:
\begin{lemma}\label{Jtrans}
The mapping~\eqref{Jconjugation} acts linearly on the vector spaces
$\myspan\{J_i^-,J_i^0,J_i^+\}$, $i=1,\dots,N$, as
\begin{equation}\label{Jtransformation}
\begin{pmatrix}
 \widehat J_i^+(\z)\\[.5mm]
 \widehat J_i^0(\z)\\[.5mm]
 \widehat J_i^-(\z)
\end{pmatrix}
=\frac{1}{\Delta}
\begin{pmatrix}
  \alpha^2\, & 2\alpha\beta\, & \beta^2
  \vphantom{\widehat J_i^+}\\[.5mm]
  \alpha\gamma\, & \alpha\delta+\beta\gamma\, & \beta\delta
  \vphantom{\widehat J_i^0}\\[.5mm]
  \gamma^2\, & 2\gamma\delta\, & \delta^2
  \vphantom{\widehat J_i^-}
\end{pmatrix}\!
\begin{pmatrix}
  J_i^+(\z)\vphantom{\widehat J_i^+}\\[.5mm]
  J_i^0(\z)\vphantom{\widehat J_i^0}\\[.5mm]
  J_i^-(\z)\vphantom{\widehat J_i^-}
\end{pmatrix}.
\end{equation}
\end{lemma}
It follows from the previous lemma that the operator
$\widehat H^*$ defined by
$$
\widehat H^*=\mu_m(\z)\,H^*\big(\w(\z)\big)\,\mu_m^{-1}(\z)
$$
is still a second degree polynomial in the Dunkl operators
$J_i^\ep(\z)$, whose coefficients $\hat c_{\ep\ep'}$, $\hat
c_{\ep}$ can be easily computed using
Eq.~\eqref{Jtransformation}. The corresponding gauge spin
Hamiltonian $\widehat\BH$ can be obtained from
Eq.~\eqref{gaugeham} by replacing the coefficients $c_{\ep\ep'}$
and $c_{\ep}$ by their counterparts $\hat c_{\ep\ep'}$ and $\hat
c_{\ep}$. In particular, the polynomials $P(z)$ and $Q(z)$ in
Eq.~\eqref{Q} are replaced by the polynomials
\begin{equation}\label{hatPQ}
  \widehat P(z)=\hat c_{++}z^4+\hat c_{0+}z^3+\hat c_{00}z^2+
  \hat c_{0-}z+\hat c_{--}\,,\qquad
  \widehat Q(z)=\hat c_{+}z^2+\hat c_{0}z+\hat c_{-}\,.
\end{equation}
Expressing the coefficients $\hat c_{\ep\ep'}$ and $\hat c_{\ep}$
in terms of the original coefficients $c_{\ep\ep'}$ and
$c_{\ep}$, one easily arrives to the explicit formulas
\begin{equation}\label{hatPQ2}
\widehat P(z)={(\gamma z + \delta)^4 \over
\Delta^2}\,P\!\left({\alpha z + \beta\over \gamma z +
\delta}\right)\,,\qquad
\widehat Q(z)={(\gamma z + \delta)^2 \over
\Delta}\,Q\!\left({\alpha z + \beta\over \gamma z +
\delta}\right)\,.
\end{equation}
Recall~\cite{Ol99} that the (irreducible) multiplier
representation $\rho_{n,i}$
of ${\rm GL}(2,\R)$ on the space of univariate polynomials
of degree at most $n$ is defined by the linear transformations
$$
p(z)\:\mapsto\:\hat p(z)=\Delta^i(\gamma
z + \delta)^n\,p\left({\alpha z + \beta\over \gamma z +
\delta}\right).
$$
By Eq.~\eqref{hatPQ2}, the polynomials $P$ and $Q$ defining $\BH$
transform according to the representations $\rho_{4,-2}$ and
$\rho_{2,-1}$, respectively. Note also that the Dunkl
operators~\eqref{Js} transform according to the representation
$\rho_{2,-1}$; see Eq.~\eqref{Jtransformation}. The (nonzero)
orbits of the representation $\rho_{4,-2}$ can be parametrized by
the following canonical forms~\cite{gur64,GKO94}:
\begin{gather*}
\pm 1\,,\quad z\,,\quad \pm\nu(1-z^2)\,,\quad \pm\nu(1+z^2)\,,\quad
\pm\nu z^2\,,\quad \pm\nu(1+z^2)^2\,,\\
\pm\nu(1-z^2)(1-k^2z^2)\,,\quad
\pm\nu(1+z^2)\big(1+k^2z^2\big)\,,\quad
\nu(1-z^2)\big(1-k^2+k^2 z^2\big)\,,
\end{gather*}
where $\nu>0$ and $0<k<1$. The above list of canonical forms can
be further reduced without any loss of generality using
the {\em complex} projective (linear) transformation $w=\iu z$. Since the
projective transformation $w=\iu z$ also induces the mapping
$c_\ep\mapsto \hat c_\ep=\iu^\ep c_\ep$ for the coefficients of
the polynomial $Q$, the resulting canonical form leads to a real
potential provided that the initial coefficients $c_\pm$ are
purely imaginary. The reduced list of canonical forms will be
conveniently taken as
\begin{equation}
\begin{aligned}
& 1)\quad 1\,,\qquad\quad & 6)&\quad \nu(1+z^2)^2\,,\\
& 2)\quad z\,,\qquad\quad & 7)&\quad \nu(1-z^2)(1-k^2z^2)\,,\\
& 3)\quad \nu(z^2-1)\,,\qquad\quad & 8)&\quad \nu(z^2-1)\big(1-k'^2z^2\big)\,,\\
& 4)\quad \nu(1-z^2)\,,\qquad\quad & 9)&\quad \nu(1-z^2)\big(k'^2+k^2 z^2\big)\,,\\
& 5)\quad \nu z^2\,,\qquad\quad &\label{canonic}
\end{aligned}
\end{equation}
where $k'^2=1-k^2$. By choosing $P(z)$ in
Eqs.~\eqref{gaugeham}, \eqref{Q}
in each of the canonical forms~\eqref{canonic}, one obtains a complete
classification of the spin CS models with potential~\eqref{pot},
which are (Q)ES by construction. Recall that in the one-particle
case the canonical forms 1), 2), 3), 5) give rise
to rational or hyperbolic potentials, while the remaining five
yield periodic (trigonometric or elliptic) potentials~\cite{GKO94}.\\

\begin{remark}\label{rem.proj}
The operator $H^*$ in~\eqref{Jcombination} is easily seen to
preserve the module $\cS_m(\z)$ of symmetric polynomials in $z_1,\dots,z_N$
of degree at most $m$, on which the permutation operators
$K_{ij}$ act as the identity. Hence the antisymmetrizer $\La$
acts on the space $\cS_m\otimes\fS$ as the tensor product
$\id\otimes\La_0$, where $\La_0$ is the spin antisymmetrization
operator, and the $H$-invariant module
$\mu\La\big(\cS_m\otimes\fS\big)$ factors as the tensor product
$(\mu\,\cS_m)\otimes(\La_0\fS)$. The spin permutation operators
$S_{ij}$ on this module reduce to $-\id$. Therefore, the
restriction of the Hamiltonian $H$ to this space is simply $
\big(H\big|_{S_{ij}\to-1}\big)\otimes\id $. Thus the {\em scalar}
Schr\"odinger operator $H\big|_{S_{ij}\to-1}$ leaves invariant
the module $\mu\,\cS_m$. It follows that replacing the spin
permutation operators $S_{ij}$ by $-1$ in one of the (Q)ES spin
potentials listed below, one obtains a corresponding (Q)ES scalar
potential. The scalar potentials so constructed include as
particular cases all the potentials presented in~\cite{GGR00}.
\end{remark}

For each of the canonical forms~\eqref{canonic}, the potential
turns out to decompose as
\begin{equation}\label{U+Vint}
V(\x) = \sum_i U(x_i)+V_{\text{int}}(\x)\,,
\end{equation}
where $U$ plays the role of a external field potential, and the
interaction potential $V_{\text{int}}$ is of the form
\begin{equation}\label{Vint}
V_{\text{int}}(\x)=\sum_{i<j}\big(V^-(x_i-x_j)+V^+(x_i+x_j)\big)\,
a(a+S_{ij})\,,
\end{equation}
with either $V^+=0$ or $V^+=V^-$. Indeed, making use of the
identities~\eqref{A7}--\eqref{A15} in the Appendix and recalling that
$c_{0+}=0$ for all the canonical forms,
the potential~\eqref{pot} reduces after some algebra
to the form~\eqref{U+Vint} with
\begin{align*}
&U=c_{++}(1-b^2)z^2+b\,c_{+}z
+{1\over 16\,P(z)}\,\Big(3P'(z)^2+4\tilde{Q}(z)^2-8\tilde{Q}(z)P'(z)\Big)\,,\\
&V_{\text{int}}=2\sum_{i<j}(z_i-z_j)^{-2}
\Big(c_{++}z_i^2z_j^2+c_{00}z_iz_j+\frac{c_{0-}}2\,(z_i+z_j)+
c_{--}\Big)\,a(a+S_{ij})\,.
\end{align*}
We have discarded here a constant term of the form~\eqref{V0}, in accordance
with the observation at the end of Section~\ref{sec.cons}. The interaction
potential takes the form~\eqref{Vint} after performing the change of
variables~\eqref{r8} and using some identities for the corresponding
function $\ze(x)$. Moreover, if the function $V^+$ is nonzero,
it can be reduced to $V^-$ by a suitable coordinate translation
(see Remark~3).

We shall now present the list of (Q)ES spin many-body potentials
obtained from the canonical forms~\eqref{canonic}.
Note that the scaling $(c_{\ep\ep'},c_\ep)\mapsto(\la c_{\ep\ep'},\la c_\ep)$
induces the mapping
\begin{align*}
V(\x\,;c_{\ep\ep'},c_{\ep})\;&\mapsto\;
V(\x\,;\la\,c_{\ep\ep'},\la\,c_{\ep})
=\la\,V(\sqrt\la\,\x\,;c_{\ep\ep'},c_{\ep})\\
\mu(\x\,;c_{\ep\ep'},c_{\ep})\;&\mapsto\;
\mu(\x\,;\la\,c_{\ep\ep'},\la\,c_{\ep})\propto
\mu(\sqrt\la\,\x\,;c_{\ep\ep'},c_{\ep})
\end{align*}
of the corresponding potentials and gauge factors.
For this reason we shall list the
potentials for a suitably chosen value of the parameter $\nu$
in Cases 3--9, or a suitable multiple of $P(z)$ in Cases 1,2.
The notation
$$
x_{ij}^\pm=x_i\pm x_j\,;\qquad
\al_\ep=\frac{c_\ep}4\,,\quad\ep=\pm,0\,;\qquad
\al=\al_++\al_0+\al_-
$$
shall be employed in what follows.\\

\ni{\bf Case 1.}\quad $\displaystyle{P(z)=\frac14}$.\\
{\em Change of variables}:\quad $\displaystyle{z=\frac x2}$.\\
{\em Gauge factor}:
$$
\mu(\x)=\prod_{i<j}(x_{ij}^-)^a
\prod_i\exp\!\bigg(\frac13\,\al_+x_i^3+\al_0x_i^2+4\al_-x_i\bigg).
$$
{\em External potential}:
$$
U(x)=\al_+^2 x^4+4\al_0 \al_{+}x^3
+4(\al_0^2+2\al_-\al_+)\,x^2+2(8\al_-\al_0+b\al_+)\,x\,.
$$
{\em Interaction potential}:
$$
V_{\text{int}}(\x)=2\sum_{i<j}(x_{ij}^-)^{-2}\,a(a+S_{ij})\,.
$$\vskip 1mm

\ni{\bf Case 2.}\quad $P(z)=4z$.\\
{\em Change of variables}:\quad $z=x^2$.\\
{\em Gauge factor}:
$$
\mu(\x)=\prod_{i<j}\big(x_{ij}^-\,x_{ij}^+\big)^a
\prod_ix_i^{\frac12(1-b)+\al_-}
\exp\!\bigg(\frac1{4}\,\al_+x_i^4+\frac12\,\al_0x_i^2\bigg).
$$
{\em External potential}:
$$
U(x)=\al_+^2x^6+2\al_0\al_+x^4+\big(\al_0^2+\al_+(3b+2\al_-)\big)x^2
+{1\over 4}\,\big((2\al_--b)^2-1\big){1\over x^2}\,.
$$
{\em Interaction potential}:
$$
V_{\text{int}}(\x)=2\sum_{i<j}\Big(\big(x_{ij}^-\big)^{-2}
+\big(x_{ij}^+\big)^{-2}\Big)\,a(a+S_{ij})\,.
$$\vskip 1mm

\ni{\bf Case 3.}\quad $P(z)=4(z^2-1)$.\\
{\em Change of variables}:\quad $z=\cosh{2x}$.\\
{\em Gauge factor}:
$$
\mu(\x)=\prod_{i<j}\!\big(\sinh x_{ij}^-\,\sinh x_{ij}^+\big)^a\!
\prod_i (\sinh x_i)^{\frac12(1+\al-b)}\\
(\cosh x_i)^{\frac12(1+2\al_0-\al-b)}
\exp\!\Big(\frac{\al_+}2\cosh 2x_i\Big).
$$
{\em External potential}:
\begin{align*}
U(x)&=\al_+^2\cosh^2 2x+2\al_+(\al_0+b)\cosh 2x
+2(\al_++\al_-)(\al_0-b)\cosh 2x\,\sinh^{-2}\!2x\\
&\quad+\big((\al_++\al_-)^2+(\al_0-b)^2-1\big)\,\sinh^{-2}\!2x\,.
\end{align*}
{\em Interaction potential}:
$$
V_{\text{int}}(\x)=2\sum_{i<j}
\Big(\sinh^{-2}\!x_{ij}^-+\sinh^{-2}\!x_{ij}^+\Big)\,a(a+S_{ij})\,.
$$\vskip 1mm

\ni{\bf Case 4.}\quad $P(z)=4(1-z^2)$.\\
{\em Change of variables}:\quad $z=\cos{2x}$.\\
{\em Gauge factor}:
$$
\mu(\x)=\prod_{i<j}\!\big(\sin x_{ij}^-\,\sin x_{ij}^+\big)^a\!
\prod_i (\sin x_i)^{\frac12(1-\al-b)}
(\cos x_i)^{\frac12(1+\al-2\al_0-b)}
\exp\!\Big(-\frac{\al_+}2\cos 2x_i\Big).
$$
{\em External potential}:
\begin{align*}
U(x)&=-\al_+^2\cos^2 2x+2\al_+(b-\al_0)\cos 2x
+2(\al_++\al_-)(b+\al_0)\cos 2x\,\sin^{-2}\!2x\\
&\quad+\big((\al_++\al_-)^2+(b+\al_0)^2-1\big)\sin^{-2}\!2x\,.
\end{align*}
{\em Interaction potential}:
$$
V_{\text{int}}(\x)=2\sum_{i<j}
\Big(\sin^{-2}\!x_{ij}^-+\sin^{-2}\!x_{ij}^+\Big)\,a(a+S_{ij})\,.
$$\vskip 1mm

\ni{\bf Case 5.}\quad $P(z)=4z^2$.\\
{\em Change of variables}:\quad $z=\e^{2x}$.\\
{\em Gauge factor}:
$$
\mu(\x)=\prod_{i<j}\sinh^a\!x_{ij}^-
\prod_i
\exp\bigg[\frac12\Big(\al_+\e^{2x_i}-\al_-\e^{-2x_i}\Big)
+(\al_0-m)x_i\bigg].
$$
{\em External potential}:
$$
U(x)=\al_+^2\e^{4x}+2\al_+(\al_0+b)\,\e^{2x}+2\al_-(\al_0-b)\,\e^{-2x}
+\al_-^2\e^{-4x}\,.
$$
{\em Interaction potential}:
$$
V_{\text{int}}(\x)=2\sum_{i<j}\sinh^{-2}\!x_{ij}^-\:a(a+S_{ij})\,.
$$\vskip 1mm

\ni{\bf Case 6.}\quad $P(z)=(1+z^2)^2$.\\
{\em Change of variables}:\quad $z=\tan x$.\\
{\em Gauge factor}:
$$
\mu(\x)=\prod_{i<j}\sin^a\!x_{ij}^-
\prod_i \cos^m\!x_i\,
\exp\Big[(\al_++\al_-)x_i+\frac12\,(\al_--\al_+)\sin 2x_i-\frac12\,\al_0\cos 2x_i\Big].
$$
{\em External potential}:
\begin{align*}
U(x)&=\frac12\big((\al_+-\al_-)^2-\al_0^2\big)\cos 4x
+2\big(\al_-^2-\al_+^2+b \al_0)\cos 2x\\
&\quad+\al_0(\al_--\al_+)\sin 4x+2\big(\al_0(\al_++\al_-)+b(\al_+-\al_-)\big)\sin 2x\,.
\end{align*}
{\em Interaction potential}:
$$
V_{\text{int}}(\x)=2\sum_{i<j}\sin^{-2}\!x_{ij}^-\:a(a+S_{ij})\,.
$$\vskip 1mm

\ni{\bf Case 7.}\quad $P(z)=4(1-z^2)(1-k^2z^2)$.\\[1mm]
{\em Change of variables}:\quad $\displaystyle z=\frac{\cn\,2x}{\dn\,2x}$.\\
{\em Gauge factor}:
\begin{align*}
\mu(\x)&=\prod_{i<j}\bigg(\frac{\sn\,x_{ij}^-\,\sn\,x_{ij}^+}
{1-k^2\,\sn^2x_{ij}^-\,\sn^2x_{ij}^+}\bigg)^{\!a}
\prod_i \bigg[
(\sn\,2x_i)^{\frac12\big(1-b-\frac\al{k'^2}\big)}
(\dn\,2x_i)^m\\
&\qquad\qquad\times
\big(\dn\,2x_i+\cn\,2x_i\big)^{\frac{\al_++\al_-}{2k'^2}}
\big(\dn\,2x_i+k\,\cn\,2x_i\big)^{-\frac{\al_++k^2\al_-}{2kk'^2}}
\bigg]\,.
\end{align*}
Here (and also in Cases 8 and 9) the functions
$\sn\,x\equiv\sn(x|k)$, $\cn\,x\equiv\cn(x|k)$, and
$\dn\,x\equiv\dn(x|k)$ are the usual Jacobian elliptic
functions of {\em modulus} $k$, and $k'=\sqrt{1-k^2}$ is the
{\em complementary modulus}.\\
{\em External potential}:
$$
U(x)=A_7\,\sn^22x+B_7\,\cn\,2x\,\dn\,2x
+\sn^{-2}2x\,(C_7+D_7\,\cn\,2x\,\dn\,2x)\,,
$$
where
\begin{align*}
& A_7=k^2(b^2-1)+\frac{k^2\al_0}{k'^2}\Big(\frac{\al_0}{k'^2}-2b\Big)
+{1\over k'^4}\,(\al_++k^2\al_-)^2\,,\\
& B_7=\frac2{k'^2}\,(\al_++k^2\al_-)\Big(b-\frac{\al_0}{k'^2}\Big)\,,\\
& C_7=b^2-1+\frac{\al_0}{k'^2}\Big(\frac{\al_0}{k'^2}+2b\Big)
+\frac 1{k'^4}\,(\al_++\al_-)^2\,,\\
& D_7=\frac2{k'^2}\,(\al_++\al_-)\Big(b+\frac{\al_0}{k'^2}\Big)\,.
\end{align*}
{\em Interaction potential}:
$$
V_{\text{int}}(\x)=2\sum_{i<j}\bigg(
{\cn^2x_{ij}^-\,\dn^2x_{ij}^-\over\sn^2x_{ij}^-}+
{\cn^2x_{ij}^+\,\dn^2x_{ij}^+\over\sn^2x_{ij}^+}\bigg)a(a+S_{ij})\,.
$$\vskip 1mm

\ni{\bf Case 8.}\quad $P(z)=4(z^2-1)(1-k'^2z^2)$.\\[1mm]
{\em Change of variables}:\quad $\displaystyle z=\frac1{\dn\,2x}$\,.\\
{\em Gauge factor}:
\begin{align*}
\mu(\x)&=\prod_{i<j}\bigg(
\frac{\sn\,x_{ij}^-\,\sn\,x_{ij}^+\,\cn\,x_{ij}^-\,\cn\,x_{ij}^+}
{1-k^2\,\sn^2x_{ij}^-\,\sn^2x_{ij}^+}\bigg)^{\!a}
\prod_i \bigg[
(\cn\,2x_i)^{\frac12\big[1-b-\frac1{k'k^2}(\al_++k'\al_0+k'^2\al_-)\big]}\\
&\hspace{1.25em}\times
(\sn\,2x_i)^{\frac12\big(1-b+\frac\al{k^2}\big)}
(\dn\,2x_i)^m
\big(1+\dn\,2x_i\big)^{-\frac{\al_++\al_-}{2k^2}}
\big(k'+\dn\,2x_i\big)^{\frac{\al_++k'^2\al_-}{2k'k^2}}
\bigg].
\end{align*}
{\em External potential}:
$$
U(x)=\sn^{-2}2x\,(A_8+B_8\,\dn\,2x)+\cn^{-2}2x\,(C_8+D_8\,\dn\,2x)\,,
$$
where
\begin{align*}
& A_8=b^2-1+\frac{\al_0}{k^2}\Big(\frac{\al_0}{k^2}-2b\Big)
+\frac 1{k^4}\,(\al_++\al_-)^2\,,\\
& B_8=\frac2{k^2}\,(\al_++\al_-)\Big(\frac{\al_0}{k^2}-b\Big)\,,\\
& C_8=k'^2(b^2-1)+\frac{k'^2\al_0}{k^2}\Big(\frac{\al_0}{k^2}+2b\Big)
+{1\over k^4}\,(\al_++k'^2\al_-)^2\,,\\
& D_8=\frac2{k^2}\,(\al_++k'^2\al_-)\Big(\frac{\al_0}{k^2}+b\Big)\,.
\end{align*}
{\em Interaction potential}:
$$
V_{\text{int}}(\x)=2\sum_{i<j}\bigg(
{\dn^2x_{ij}^-\over\sn^2x_{ij}^-\,\cn^2x_{ij}^-}
+{\dn^2x_{ij}^+\over\sn^2x_{ij}^+\,\cn^2x_{ij}^+}\bigg)a(a+S_{ij})\,.
$$\vskip 1mm

\ni{\bf Case 9.}\quad $P(z)=4(1-z^2)(k'^2+k^2z^2)$.\\
{\em Change of variables}:\quad $z=\cn\,2x$.\\
{\em Gauge factor}:
\begin{align*}
\mu(\x)&=\prod_{i<j}\bigg(\frac{\sn\,x_{ij}^-\,\sn\,x_{ij}^+
\,\dn\,x_{ij}^-\,\dn\,x_{ij}^+}
{1-k^2\sn^2x_{ij}^-\,\sn^2x_{ij}^+}\bigg)^{\!a}
\prod_i \Bigg[(\sn\,2x_i)^{\frac12(1-\al-b)}
(\dn\,2x_i)^{\frac12(1+\al_0-b)}\\
&\qquad\qquad\times
\big(1+\cn\,2x_i\big)^{\!\frac12(\al_++\al_-)}
\exp\!\bigg[\frac{k^2\al_--k'^2\al_+}{2kk'}\,
\tan^{-1}\!\Big(\frac k{k'}\,\cn\,2x_i\Big)\bigg]\Bigg]\,.
\end{align*}
{\em External potential}:
$$
U(x)=\dn^{-2}2x\,(A_9+B_9\,\cn\,2x)+\sn^{-2}2x\,(C_9+D_9\,\cn\,2x)\,,
$$
where
\begin{align*}
& A_9=k'^2(1-b^2)+k'^2\al_0(2b-\al_0)+
{1\over k^2}\,\big(k'^2\al_+-k^2\al_-\big)^2\,,\\
& B_9=2(b-\al_0)\big(k'^2\al_+-k^2\al_-\big)\,,\\
& C_9=(b+\al_0)^2+(\al_++\al_-)^2-1\,,\\
& D_9=2(b+\al_0)(\al_++\al_-)\,.
\end{align*}
{\em Interaction potential}:
$$
V_{\text{int}}(\x)=2\sum_{i<j}\bigg(
{\cn^2x_{ij}^-\over\sn^2x_{ij}^-\,\dn^2x_{ij}^-}
+{\cn^2x_{ij}^+\over\sn^2x_{ij}^+\,\dn^2x_{ij}^+}
\bigg)a(a+S_{ij})\,.
$$\vskip 1mm

\begin{remark}
In Case~7, the alternative canonical form
$$
P(z)=4z^3-g_2z-g_3\,,\qquad g_2^3>27g_3^2\,,
$$
leads to a spin generalization of the QES potential involving Weierstrass
functions studied in~\cite{GGRprep}. The corresponding change of variables
is $z=\wp(x+\om_3)$, where $\wp(x)=\wp(x|g_2,g_3)$ is the Weierstrass
function with invariants $g_2$, $g_3$, and
$2\om_3$ is its purely imaginary fundamental period. The gauge factor reads
\begin{align*}
\mu(x)&=\prod_{i<j}\big(\wp(x_i+\omega_3)-\wp(x_j+\omega_3)\big)^a
\prod_i\bigg[\big(\wp'(x_i+\omega_3)\big)^\be\\
&\qquad\times\big(\wp(x_i+\omega_3)-e_1\big)^{\ga_1}
\big(\wp(x_i+\omega_3)-e_2\big)^{\ga_2}
\big(\wp(x_i+\omega_3)-e_3\big)^{\ga_3}\bigg],
\end{align*}
where
\begin{align*}
& \be=\frac12\,(1-b)+\frac{\al_+}3\,,\\[1mm]
& \ga_j=\frac{g_2\al_++12(e_j\al_0+\al_-)}{24(e_j-e_k)(e_j-e_l)}\,,
\qquad (j,k,l)=\text{cyclic permutation of }(1,2,3)\,,
\end{align*}
and $e_j$ are the real (different) roots of $P(z)$. The external
and interaction potentials are given by
$$
U(x)=4\be(\be-1)\wp(2x)+A\wp(x+\omega_3)+\big(B\wp^2(x+\omega_3)
+C\wp(x+\omega_3)+D\big)\big(\wp'(x+\omega_3)\big)^{-2},
$$
where
\begin{align*}
& A=2\al_0+\frac49\,\al_+(2\al_++3b)\,,\\
& B=4\al_0^2+\frac13\,(2\al_+-3b)(12\al_-+g_2\al_+)\,,\\
& C=2\al_0\big(4\al_-+g_2(\al_+-b)\big)\,,\\
& D=\Big(g_2b+4\al_--\frac13\,g_2\al_+\Big)\Big(\al_-+\frac1{12}\,g_2\al_+\Big)
+g_3\al_0(3b-2\al_+)\,,
\end{align*}
and
$$
V_{\rm int}(\x)=2\sum_{i<j}\big(\wp(x_{ij}^-)+\wp(x_{ij}^+)\big)\,a(a+S_{ij})\,.
$$
The restriction of this spin model
to the polynomial space $\cS_m$ (see Remark~4)
yields the scalar elliptic CS model in~\cite{GGRprep}
provided $\al_0=0$ and $\al_-=-g_2\al_+/12$.
\end{remark}

\begin{remark}
Case 1 with $\al_+=0$ yields the rational Calogero $A_N$ spin model.
Case 5 with $\al_+\al_-=0$ is the model studied by Inozemtsev~\cite{In97},
while for $\al_+=\al_-=0$ the hyperbolic Sutherland $A_N$ spin model is obtained.
Case 6 with $\al_+-\al_-=\al_0=0$ is the trigonometric Sutherland $A_N$ spin
model. The remaining potentials are new.

In Cases 1--5, the potential is ES if $\al_+=0$. In Case 5, the potential
is also ES for $\al_-=0$. The only ES potential in Case 6 is the trigonometric
Sutherland potential ($\al_+-\al_-=\al_0=0$). The remaining potentials,
including all the elliptic potentials in Cases 7--9, are QES.
\end{remark}

\begin{remark}
In order to qualify as a physical wavefunction, the spin
functions $\psi(\x)$ constructed in Section~\ref{sec.cons}
must vanish at all points in which the
corresponding potential $V(\x)$ is singular. In addition, in
Cases~1,2,3,5 the function $\psi$ is required to be square-integrable
over a suitable domain of $\R^N$. Both requirements impose certain constraints
on the parameters $\al_\ep$ and $a$ defining the potential.
A detailed analysis of the necessary and sufficient
conditions on these parameters lies beyond the scope of this paper;
see~\cite{GKO93} for a complete solution of the
analogous one-particle problem. However, it is not difficult
in each case to provide sufficient conditions for the above requirements
to hold. For example, in Case~5 the spin functions $\psi$ vanish
at the singularities of the
potential if and only if $a>0$, and are square-integrable over
the domain $\{\x\in\R^N\::\:x_1>\dots>x_N\}$ {\em provided}
$\al_+<0$ and $\al_->0$.
\end{remark}

\begin{remark}
In Cases 1--6, the gauge factor is of the form
\begin{equation}\label{factor}
\mu(\x)=\prod_{i<j}\big[f(x_{ij}^-)\,g(x_{ij}^+)\big]^a
\prod_i h(x_i)\,,
\end{equation}
with $g=1$ or $g=f$. On the other hand, in the elliptic Cases 7--9 the gauge
factor does not factorize as~\eqref{factor}. This is consistent with
the analogous result proved by Calogero for elliptic potentials
in the scalar case~\cite{Ca75}.
\end{remark}

\section{Conclusions}

We have developed in this paper a systematic method for constructing
new families of exactly and quasi-exactly solvable spin Calogero--Sutherland
models. The key idea consists in relating the physical spin
Hamiltonian to a general quadratic combination involving the usual
$A_N$-type Dunkl operators~\eqref{r2}, \eqref{r01}, and the new
family of Dunkl operators $J_i^+$ introduced
in Section~\ref{sec.dunkl} (Eq.~\eqref{Js}).
Our approach goes beyond the Lie algebraic method extensively used
in the scalar case, since the Dunkl operators~\eqref{Js} do not span
a Lie algebra. However, they are invariant under the projective
action of ${\rm GL}(2,\R)$, a fact that is exploited in
Section~\ref{sec.class} to classify all the potentials obtained
by this method up to translations.

The potentials constructed in this paper are all invariant under the
$A_N$ group consisting of permutations of the particles' coordinates
and spins. A remarkable feature of the potentials in Cases 2--4
and 7--9 is their additional invariance under a change of sign of
the spatial coordinate of any particle. Therefore,
although these potentials are not invariant under the
full $B_N$ group of permutations and sign reversal of the
particles' coordinates {\em and} spins, they are invariant
under the restriction of the action of this group to the
spatial coordinates. These models thus occupy an intermediate
position between the usual spin CS models of $A_N$ type and the
fully $B_N$-invariant rational and trigonometric spin CS models
introduced by Yamamoto~\cite{yam95}. In fact, while Dunkl has recently
proved the exact solvability of the rational Yamamoto model~\cite{dunkl98},
there are no exact results for the eigenfunctions of its
trigonometric counterpart. It is to be expected that a suitable
extension of the method developed in this paper to the $B_N$ case
will yield new families of (Q)ES spin CS models of $B_N$ type,
including the trigonometric Yamamoto model.

\section{Appendix}

The formulae for the sums of the squares and the anticommutators
of the Dunkl operators~\eqref{Js} are given by
\begin{align}
&\sum_i(J_i^-)^2=\sum_i\pa_{z_i}^2
+2a\sum\limits_{i<j}{1\over z_i-z_j}\,
\big(\pa_{z_i}-\pa_{z_j}\big)
-2a\sum\limits_{i<j}{1\over(z_i-z_j)^2}\,(1-K_{ij}),\tag{A1}
\label{A1}\\
&\begin{aligned}
&\sum_i(J_i^0)^2=\sum_i\big(z_i^2\pa_{z_i}^2
+(2-b)\,z_i\pa_{z_i}\big)
+2a\sum\limits_{i<j}{1\over z_i-z_j}\,
\big(z_i^2\pa_{z_i}-z_j^2\pa_{z_j}\big)\\
&\qquad\qquad\quad
-a\sum\limits_{i<j}{z_i^2 + z_j^2\over (z_i-z_j)^2}\,(1-K_{ij})
+a\sum\limits_{i<j}(1-K_{ij})
+\frac{a^2}{12}{\sum_{i,j,k}}'(1-K_{ij}K_{jk})
+\frac{N m^2}4,
\end{aligned}\tag{A2}\label{A2}\\
&\begin{aligned}
&\sum_i(J_i^+)^2=\sum_j\big(z_i^4\pa_{z_i}^2
+2(2-b)\,z_i^3\pa_{z_i}\big)
+2a\sum\limits_{i<j}{1\over z_i-z_j}\,
\big(z_i^4\pa_{z_i}-z_j^4\pa_{z_j}\big)\\
&\qquad\qquad\quad
-a\sum\limits_{i<j}{z_i^4+z_j^4\over(z_i-z_j)^2}\,(1-K_{ij})
+a\sum\limits_{i<j}(z_i+z_j)^2(1-K_{ij})\\
&\qquad\qquad\quad-2am\sum\limits_{i<j}z_iz_j+m(m-1)\sum_i z_i^2,
\end{aligned}\tag{A3}\label{A3}\\
&\begin{aligned}
&\frac12\sum_i\{J_i^0,\ J_i^-\}=
\sum_i\big(z_i\pa_{z_i}^2+
\frac12(2-b)\,\pa_{z_i}\big)
+2a\sum\limits_{i<j}{1\over z_i-z_j}\,\big(z_i\pa_{z_i}-z_j\pa_{z_j}\big)\\
&\qquad\qquad\qquad\quad\;
-a\sum\limits_{i<j}{z_i+z_j\over(z_i-z_j)^2}\,(1-K_{ij}),
\end{aligned}\tag{A4}\label{A4}\\
&\begin{aligned}
&\frac12\sum_i\{J_i^+,J_i^-\}=
\sum_i\big(z_i^2\pa_{z_i}^2+(2-b)\,z_i\pa_{z_i}\big)
+2a\sum\limits_{i<j}{1\over z_i-z_j}\,\big(z_i^2\pa_{z_i}-z_j^2\pa_{z_j}\big)\\
&\qquad\qquad\qquad\quad\;
-a\sum\limits_{i<j}{z_i^2+z_j^2\over(z_i-z_j)^2}\,(1-K_{ij})
-\frac{a^2}{6}{\sum\limits_{i,j,k}}'(1-K_{ij}K_{jk})\\
&\qquad\qquad\qquad\quad\;-\frac{mN}{2}\big(1+a(N-1)\big),
\end{aligned}\tag{A5}\label{A5}\\
&\begin{aligned}
&\frac12\sum_i\{J_i^0,J_i^+\}=
\sum_i\,\big(z_i^3\pa_{z_i}^2+\frac32(2-b)\,z_i^2\pa_{z_i}\big)
+2a\sum\limits_{i<j}{1\over z_i-z_j}\,\big(z_i^3\pa_{z_i}-z_j^3\pa_{z_j}\big)\\
&\qquad\qquad\qquad\quad\;
-a\sum\limits_{i<j}{z_i^3+z_j^3\over (z_i-z_j)^2}\,(1-K_{ij})
+a\sum\limits_{i<j}(z_i+z_j)(1-K_{ij})\\
&\qquad\qquad\qquad\quad\;
+\frac{m(2m-b)}2\sum_i z_i,
\end{aligned}\tag{A6}\label{A6}
\end{align}
where
$$
b=1+m+a(N-1).
$$
Here the symbol ${\sum\limits_{i,j,k}}'$ stands for summation in
$i,j,k$ with $i\neq j\neq k\neq i$. The following identities
are needed for the computation of the potential:
\begin{align}
&{\sum\limits_{i,j,k}}'{1\over (z_i-z_j)(z_i-z_k)}=0,\tag{A7}\label{A7}\\
&{\sum\limits_{i,j,k}}'{z_i\over (z_i-z_j)(z_i-z_k)}=0,\tag{A8}\label{A8}\\
&{\sum\limits_{i,j,k}}'{z_i^2\over (z_i-z_j)(z_i-z_k)}
={1\over 3}N(N-1)(N-2),\tag{A9}\label{A9}\\
&{\sum\limits_{i,j,k}}'{z_i^3\over(z_i-z_j)(z_i-z_k)}
=(N-1)(N-2)\sum_i z_i,\tag{A10}\label{A10}\\
&{\sum\limits_{i,j,k}}'{z_i^4\over (z_i-z_j)(z_i-z_k)}
=(N-2)\sum_{i<j} (z_i+z_j)^2,\tag{A11}\label{A11}\\
&\sum_{i\neq j}{1\over z_i-z_j}=0,\tag{A12}\label{A12}\\
&\sum_{i\neq j}{z_i\over z_i-z_j}={1\over 2}N(N-1),\tag{A13}\label{A13}\\
&\sum_{i\neq j}{z_i^2\over z_i-z_j}=(N-1)\sum_i z_i,\tag{A14}\label{A14}\\
&\sum_{i\neq j}{z_i^3\over z_i-z_j}=(N-1)\sum_i z_i^2+\sum_{i<j}
z_iz_j.\tag{A15}\label{A15}
\end{align}

\begin{acknowledgements}
This work was partially supported by the DGES under grant PB98-0821.
R. Zhdanov would like to acknowledge the financial support
by the Spanish Ministry of Education and Culture during his
stay at the Universidad Complutense de Madrid.
\end{acknowledgements}

\end{document}